\documentclass[12pt,preprint]{aastex}

\shorttitle{INTEGRAL IGR J18135-1751=HESS J1813-178}
\shortauthors{Ubertini et al.}

\begin{document}


\title{INTEGRAL IGR J18135-1751=HESS J1813-178: A new cosmic high energy accelerator from keV to TeV
\altaffilmark{1}}

\author{P. Ubertini\altaffilmark{2}, L. Bassani\altaffilmark{3}, A.
Malizia\altaffilmark{3}, A. Bazzano\altaffilmark{2},
A.J.Bird\altaffilmark{4}, A.J. Dean\altaffilmark{4},   A. De
Rosa\altaffilmark{2}, F. Lebrun\altaffilmark{5}, L.
Moran\altaffilmark{4}, M. Renaud\altaffilmark{5}, J.B.
Stephen\altaffilmark{3}, R. Terrier\altaffilmark{5}, and R.
Walter\altaffilmark{6} }

 \altaffiltext{1}{Based on observations with {\it INTEGRAL},
an ESA project with instruments and science data centre funded by
ESA member states (especially the PI countries: Denmark
France,Germany, Italy, Switzerland, Spain), Czech Republic and
Poland and with the participation of Russia and the USA.}
\altaffiltext{2}{IASF/INAF, Rome, Italy, Via Fosso del Cavaliere
100, I-00133 Rome, Italy} \altaffiltext{3}{IASF/INAF, Bologna,
Italy, Via Gobetti 101, I-40129 Bologna, Italy}
\altaffiltext{4}{School of Physics and Astronomy, University of
Southampton, Highfield, Southampton, SO 17 1BJ, UK.}

\altaffiltext{5}{Sap-CEA, Saclay, F-91191 Gif-sur-Yvette, France}
\altaffiltext{6}{{\it INTEGRAL} Science Data Centre,Chemin
d'Ecogia 16,1291, Versoix, Switzerland}


\begin{abstract}

We report the discovery of a soft gamma ray source, namely IGR
J18135-1751, detected with the IBIS imager on board the INTEGRAL
satellite. The source is persistent and has a 20-100 keV
luminosity of $\sim$5.7 $\times$ 10$^{34}$ erg s$^{-1}$ (assuming
a distance of 4kpc). This source is coincident with one of the eight
unidentified objects recently reported by the HESS collaboration as part of the
first TeV survey of the inner part of the Galaxy.
Two of these new sources found along the Galactic plane, HESS J1813-178 and HESS J1614-518, have no obvious lower energy counterpart, a fact that motivated the suggestion that they might be dark cosmic ray accelerators.
HESS J1813-178 has a
strongly absorbed X-ray counterpart, the ASCA source
AGPS273.4-17.8, showing a power law spectrum with photon index
$\sim$ 1.8 and a total (galactic plus intrinsic) absorption
corresponding to N$_H$ $\sim$5 $\times$ 10$^{22}$ cm$^{-2}$.
We hypothesize that the source is a pulsar wind nebula embedded in
its supernova remnant.  The lack of X/gamma-ray
variability, the radio morphology and the ASCA spectrum are all
compatible with this interpretation. In any case we
rule out the hypothesis that HESS J1813-178 belongs to a new class
of TeV objects or that it is a cosmic "dark particle" accelerator.

\end{abstract}

\keywords{gamma rays: observations --- X-rays: observations --- X-rays:
individual (\objectname{HESS J1813-178}, \object{IGR J18135-1751},
\objectname{AGPS273.4-17.8})}

\section{Introduction}
The HESS (High Energy Stereoscopic System) collaboration has
recently reported results from the first sensitive TeV survey of the
inner part of our Galaxy \cite{aha05a}. This survey revealed the
existence of a population of high energy gamma-ray objects, most
of which were previously unknown. These findings have an important
astrophysical meaning because of the possibility of studying
cosmic particle accelerators via TeV observations. Various types
of sources in the galaxy can act as cosmic accelerators: pulsars
and their pulsar wind nebulae (PWN), supernova remnants (SNR),
star forming regions and possibly binary systems with a collapsed
object like a microquasar or a pulsar. At least for two sources of
the HESS survey (HESS J1813-178 and HESS J1614-518), no known
radio or X-ray counterparts have been found so far, raising the
possibility that they are a new class of objects. The lack of
X-ray emission, very recently confirmed by a CHANDRA observation 
for HESS J1303-631 \cite{muk05} is particularly interesting
since it suggests that the accelerated particles are nucleons
rather than high energy electrons. Therefore, the detection of X
and/or gamma-ray emission from these TeV sources is a key issue to
disentangle the mechanisms active in the different emitting
regions and, in turn, the source nature. To perform this task, the
IBIS gamma-ray imager on board INTEGRAL is a powerful tool: it
allows source detection above 20 keV and up to the MeV range (i.e.
in the "non thermal process" region) with a mCrab ($\sim $10$^{-11}$
erg cm$^{-2}$ s$^{-1}$) sensitivity in well exposed regions, an
angular resolution of 12 arcmin and a point source location
accuracy of 1-2 arcmin for moderately bright sources \cite{ube03}.
Furthermore, INTEGRAL has regularly observed the entire Galactic
Plane during the first two and half years in orbit providing,
above 20 keV the first galactic survey with unprecedented
sensitivity, i.e. of the order of a mCrab for 1 Ms exposure
\cite{bir04}. A second catalogue, utilising more sky coverage and
deeper exposures, is now in the final stages of preparation
\cite{bir05}. The first survey indicates clear associations for a
number of TeV detected objects: AX J1838.0-0655 \cite{aha05a}, SGR
A$^{\bigstar}$ \cite{aha04a}, MSH 15-52/PSR1509-58 \cite{aha05b}
and the Crab Nebula \cite{aha04b}. In this paper we report the
discovery of a newly detected IBIS/ISGRI source, namely IGR
J18135-1751, that is the soft gamma-ray counterpart of the TeV
source HESS J1813-178, one of the two objects for which lack of
detection in the X-ray/radio band suggested it might be a "dark
particle" accelerator.

\section{The HESS source and its counterparts}

HESS J1813-178 is one of the 8 previously unknown sources
found in the HESS survey
of the inner regions of the galactic plane. It is located at
R.A.(2000)=18h 13m 37.9s and Dec(2000)=-17$^{\circ}$ 50' 34'' with
a positional uncertainty in the range of 1-2'. The source does not
seem point-like although it is only slightly extended (3'), if
compared to the HESS point spread function. The statistical
significance of the TeV detection is around 9 sigma. The source is
fairly bright above 200 GeV with a flux of 12 $\times$ 10$^{-12}$
photons cm$^{-2}$ s$^{-1}$. No obvious counterparts were
found within the source extension, although HESS J1813-178 lies
close (10') to the radio source W33, a bright star forming region
characterized by a visually obscured compact radio core
(G12.8-0.2) located inside a molecular cloud complex \cite{has83} 
(see section 2.3 for a detailed discussion).

\subsection{The INTEGRAL/IBIS source}

\begin{figure}
\plotone{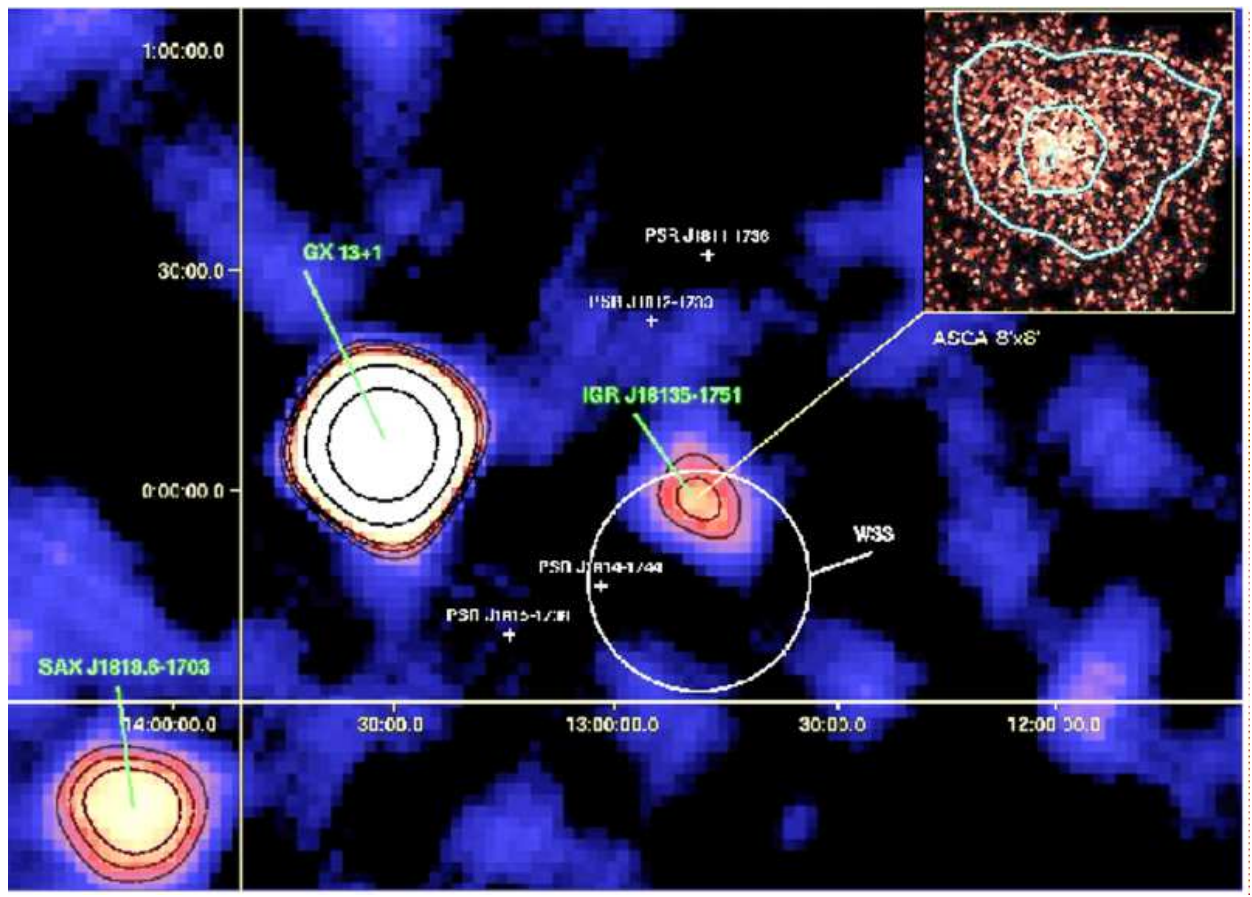} \caption{The IBIS/ISGRI 20-40 keV significance
map showing the location of IGR J18135-1751 and relative
significance contours. The gamma-ray contours shown are 6 (for the
external one), 8, 10, 20 and 40 sigma; the source spatial profile is 
compatible with the detector response to a point source.  
The extension of HESS J1813-178 as well as
the position of AGPS273.4-17.8 are both contained within the
internal IBIS/ISGRI contour. Also shown are the location (and
extension) of W33 and the 4 nearest radio pulsars (PSR
J1814-1744, PSR J1812-1733, PSR J1815-1738 and PSR J1811-1736). The
ASCA-SIS image is shown as an insert on the top right side  of
the figure. The box covers an 8'x8' region centred on the ASCA 
source position; the contour levels (1, 2 and 3 counts/pixel) provide 
marginal evidence of extended emission. 
GX13+1 and the transient source SAX J1818.6-1703 also are visible
in the image, but contribute no contamination to region around 
IGR J18135-1751, (see text for details). 
The coordinates are displayed in the galactic system.
 \label{fig1}}
\end{figure}

The sky region containing HESS J1813-178, although covered in the
first IBIS/ISGRI survey, did not provide any significant detection
due to the limited exposure dedicated to this area.  The
second survey \cite{bir05} represents a major improvement both in
exposure time and sky coverage and makes the search for a high
energy counterpart possible.  The IBIS coded mask instrument
\cite{ube03} on board INTEGRAL \cite{win03a} is made by the
combination of two detector layers: ISGRI \cite{leb03}, an upper
CdTe layer sensitive in the range 20 keV to 1 MeV and PICsIT
\cite{coc03} a bottom CsI layer sensitive in the range 200 keV to
8 MeV. In the present paper, we refer to data collected with the
low energy ISGRI detector, as the source is only detected at low
energy, namely in the range 20-100 keV. Data reported here belong
to the Core Program (i.e. were collected as part of the INTEGRAL
Galactic Plane Survey and Galactic Center Deep Exposure
\cite{win03b}) as well as to public Open Time observations, and
span from revolution 46 (February 2003) to revolution 210 (March
2004) included.  A detailed description of the source extraction
criteria can be found in Bird et al. (2004, 2005): briefly, ISGRI
images for each available pointing are generated in ten narrow
energy bands using the INTEGRAL Science Data Center (ISDC) offline
scientific analysis software OSA version 4.1 \cite{gol03},
including background uniformity corrections \cite{ter03}. Source
ghosts are removed from each image using a catalogue of sources
built iteratively and containing at the end all detected objects.
The $\sim$ 7000 images are then mosaiced using a custom tool to
produce deep all-sky maps; finally images from adjacent energy
bands are added together to obtain a map in a given energy range,
which is then used for peak detection. The primary search tool
used is SExtractor \cite{ber96} employing a simple Gaussian point
spread function filter.  Sources are then searched for in various
energy bands above a given, quite conservative,  sensitivity
threshold (typically 6 sigma) and then included in the IBIS/ISGRI
catalogue list. As a further check the map is visually inspected
to confirm detection and avoid spurious excesses due to imperfect
image cleaning. Fig. \ref{fig1} shows the 20-40 keV band image of
the region surrounding HESS J1813-178: coincident with the TeV
object an IBIS/ISGRI source (IGR J18135-1751) is detected with a
significance exceeding 10$\sigma$ at R.A.(2000)= 18h 13m 27.12s
and Dec(2000)= -17$^{\circ}$ 50' 56'' and with a positional
uncertainty of $\sim$2', corresponding to 90\% error for a 10 sigma source \cite{gro03}. Apart from the source of
interest here, two other bright objects are visible in the map:
the low mass X-ray binary GX 13+1 and the transient source SAX
J1818.6-1703, approximately 44' and  87' away from the new
IBIS/ISGRI source. Contamination from the bright source GX13+1 is
unlikely as IGR J18135-1751 is well outside the major peak of the
PSF and does not appear to be coincident with any systematic
structure from that source. Other structures present in the region
containing the source have been carefully checked in terms of
statistical significance and  PSF: they are all below the
threshold to be possibly considered real sources and do not fit
with the expected IBIS detector spatial response \cite{gol03}.

The HESS J1813-178 source is inside the first ISGRI contour  
indicating a clear spatial association between the IBIS/ISGRI 
object and the TeV source, while the formal distance
between the two sources is less than 2.2 arcmin. The mean count
rate of IGR J18135-1751 is 0.139$\pm$0.015 and 0.093 $\pm$0.016
counts s$^{-1}$ in the 20-40 and 40-100 keV bands respectively;
these count rates correspond to a flux of 1.3 and 2.1 mCrab
respectively or to an average flux of 2.1$\times$10$^{-11}$ erg
cm$^{-2} s^{-1}$ in the 20 - 100 keV band. The flux relative to
each individual pointing has been used to generate the source
light curves in different energy bands.  No burst or flares are
visible in the light curves, nor does the source show any sign of
variability; unfortunately the low significance of the detection
prevents the search for pulsations in the IBIS/ISGRI data.

\subsection{AGPS273.4-17.8, the X-ray counterpart}
We have also searched the HEASARC archive
(www.heasarc.gsfc.nasa.gov) for the presence of the source in the
data of past X-ray missions. HESS J1813-178 is located in the
galactic plane, so that it is likely that its X-ray counterpart is
heavily absorbed. The Galactic HI column density in the source
direction is 1.89 $\times$ 10$^{22}$ cm$^{-2}$ \cite{dic90}; this
value implies strong depletion of soft X-rays, more so if
molecular hydrogen is added to the column density estimation. It
is therefore likely that any soft X-ray emission is either very
weak or undetected. There is no ROSAT source inside the HESS error
circle, although a WGACAT object (1WGA J1813.7-1755) is only 5.4'
from the TeV position; this X-ray source is associated with the 10.8
V magnitude star, HD166981, of spectral type B8. Despite being
very close, this soft X-ray source is outside both the IBIS/ISGRI
error circle and the HESS extension. Similarly other soft X-ray
emitters are all located further away (more than 10' from the TeV
source) and so are incompatible with the HESS/INTEGRAL positional
uncertainty. At higher energies, we found a possible counterpart
in the ASCA archive data: AGPS273.4-17.8, which is clearly
detected by both the SIS and GIS instruments. The sky region
surrounding this X-ray source is strongly contaminated
(particularly in the GIS image) by the presence of a very bright
nearby object (likely GX13+1) and this is probably the reason why
AGPS273.4-17.8 does not appear in the ASCA galactic plane survey
\cite{sug01} despite its detection; for this reason we consider
here only the SIS data. The ASCA position is set at R.A.(2000) =
18h 13m 35.8s and Dec(2000) = -17$^{\circ}$ 49' 43.35'' with an
associated uncertainty of 1' in radius and is therefore contained
within the first IBIS/ISGRI contour. In the X-ray band the source
is fairly bright showing a 2-10 keV flux (corrected for
absorption) of 1.8 $\times$ 10$^{-11}$ erg cm$^{-2}$ s$^{-1}$; the
broad band spectrum (after background subtraction and correction)
is characterized by an absorbed power law with a hard photon index
$\Gamma$=1.76$\pm0.25$ and a column density
N$_{H}$=5.15$^{+1.70}_{-1.24}$ $\times$ 10$^{22}$ cm$^{-2}$ 
in excess of the galactic value (errors correspond to 90\%
confidence level for a single parameter variation, i.e.
$\Delta\chi$$^{2}$=2.7, see Xspec V11.3 manual). On top of this
power law, the source shows marginal evidence of an extended soft
emission (see insert in Fig. \ref{fig1}). The best fit to this
component is obtained with a thermal Bremsstrahlung with
kT=0.42$^{+0.30}_{-0.14}$ keV, absorbed by a column density
compatible with the galactic value. However, the total emitted
power of this possible soft excess, not detected by Brogan et al. (2005), is irrelevant being 50-100
times less than the synchrotron power law component. The SIS light
curve shows no variability over the 100ks observation period and
the minimum data binning of 300 s prevents the search for fast
pulsation. Detection by ASCA and INTEGRAL over different observing
periods suggests that the source is a
persistent rather than a transient object.\\

\subsection{Counterparts at other wavelengths}

\begin{figure}
\plotone{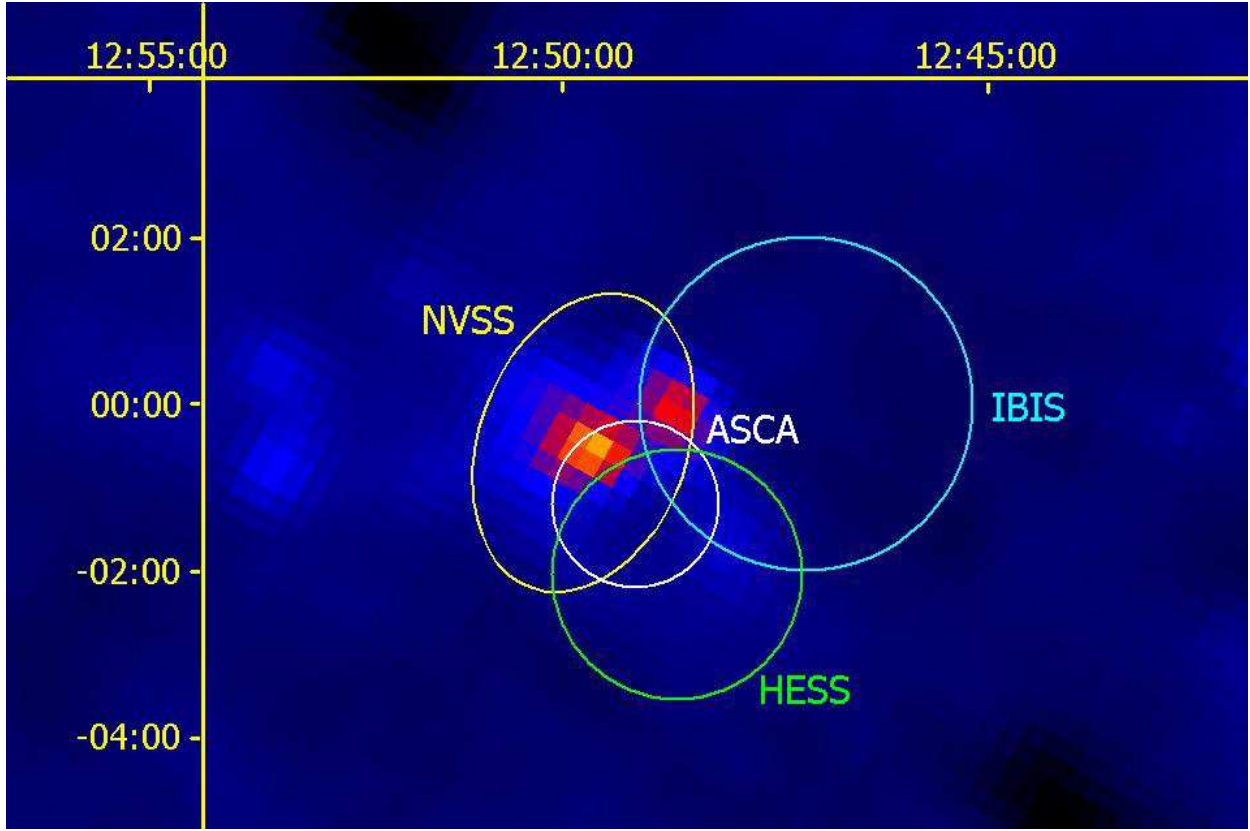} \caption{NVSS (20 cm) image of the region
surrounding IGR J18135-1751/HESS 1813-178. 
The ASCA, IBIS/ISGRI, and HESS sources (with
radii of 1', 2' and 1.5' respectively), are superimposed as is 
the extent of the NVSS J181334-17849 (ellipse).  
The radio resolution is 45'' FWHM and the rms noise is 0.45
mJy/beam. The coordinates are displayed in the galactic system.\label{fig2}}
\end{figure}

The column density measured in X-rays (galactic plus intrinsic)
implies a visual extinction Av$\sim$27 \cite{pre95}, sufficient to
obscure any optical counterpart. Therefore searches for candidate
objects at shorter frequencies should concentrate on data less
affected by absorption, such as in the radio waveband. Indeed, we
find a bright NVSS (NRAO VLA Sky Survey, \cite{con98}), radio
source within the ASCA positional uncertainty: NVSS J181334-174849
with coordinates R.A.(2000)=18h 13m 34.32s and
Dec(2000)=-17$^{\circ}$ 48' 49.1'' (see Fig. \ref{fig2}). This
source has a 20 cm flux of 327 mJy, a positional uncertainty of
only 1" and it is not polarized; the source axes are 1.85' and
1.25' (ellipse in Fig. \ref{fig2}), indicating that the object is
probably extended at 20 cm. It is located in a crowded region
showing 3 other objects, all of which have much lower flux at 20
cm. This radio complex is contained within both the IBIS/HESS
circles, but only the bright object, NVSS  J181334-174849, is
possibly associated to the ASCA source. Within its 1" positional
uncertainty, we do not find any optical and/or infrared
counterpart. 
Recently, Brogan et al. (2005) have discovered non--thermal radio 
emission from a young shell type SNR (G12.8-0.02)
associated with the radio complex NVSS. 
The source is reported as extended (3'), located at a distance $\ge$ 4
kpc and is positional coincident with HESS J1813-178. Helfand et
al. (2005) conclude the TeV emission is due to inverse Compton
scattering of starlight from W33 off the high energy electrons from
G12.8-0.02 responsible for synchrotron X-Ray emission. 

Finally, there are 4 radio pulsars within 30' of HESS J1813-178, 
but despite being
potential candidates, their distances from HESS J1813-178 argue
against a possible association. Clearly a search for likely
candidates inside the HESS source extension is the best way to
approach the problem of source identification. This is possible
via archive data as well as through dedicated observations such as
those performed by INTEGRAL.

\section{Discussion}

The basic evidence of the present work is that HESS J1813-178 has
an X-ray counterpart with a power law emission  from 2
to 10 keV, a coincident  soft gamma-ray source emitting up to 100
keV and an associated radio counterpart. The presence of a
spatially-coincident X to soft gamma-ray flux, with a power law
type emission process, rules out the hypothesis that this source
belongs to a "new class" of galactic objects or that it is a cosmic
"dark particle" accelerator: it is a non-thermal source,
accelerating electrons and positrons which radiate through
synchrotron and inverse Compton mechanism. This is suggestive of
the presence of a PWN/SNR, as already found in most newly detected
TeV objects (Aharonian et al. 2005a, b and references therein).
On the other hand, IBIS/ISGRI preferentially sees plerionic type
supernovae: 6 out of 7 remnants detected in the second catalogue
are associated with pulsar wind nebulae and only one (CasA) has a
shell type morphology \cite{bir05}. Our findings are in agreement 
with the recent radio results of
Brogan et al. (2005) and Helfand et al. (2005); 
this indicates that indeed
we are dealing with a shell type supernova remnant. However, it
cannot be ruled out at this stage that this supernova is of
composite type, e.g. a centrally filled morphology embedded inside
a shell like remnant. In these objects, the central radio/X-ray
emission is interpreted as synchrotron radiation from a bubble of
relativistic particles supplied by the central source which is
generally an isolated spinning down pulsar like PSR 1509-58.
Unfortunately, the
weakness of the gamma-ray signal and the integration time of the
X-ray data do not permit further investigation of a possible
pulsed emission.

On the other hand, the binary scenario, e.g. a typology similar to
HESS/PSR B1259-63, where the gamma-rays are generated by inverse
Compton scattering of electrons at the termination shock region of
the pulsar wind of the companion star photons, is unlikely. Apart
from the radio results, these systems have variable emission which
is not present in our data. In fact, a long integration time is
necessary to detect the gamma-ray source in view of its weakness
and the ASCA X-ray data are compatible with no flux variation over
the total observing time of about 100 ks. The good spectral match
between INTEGRAL and ASCA data can be taken as further evidence
for a constant X/gamma-ray  emission.\\
Finally, the scenario proposed by Helfand et al. (2005) where the
TeV source is due to the up-scattering of starlight from W33 by
electrons accelerated in the SNR, is not incompatible with the
IBIS/ISGRI detection.

\begin{figure}
\plotone{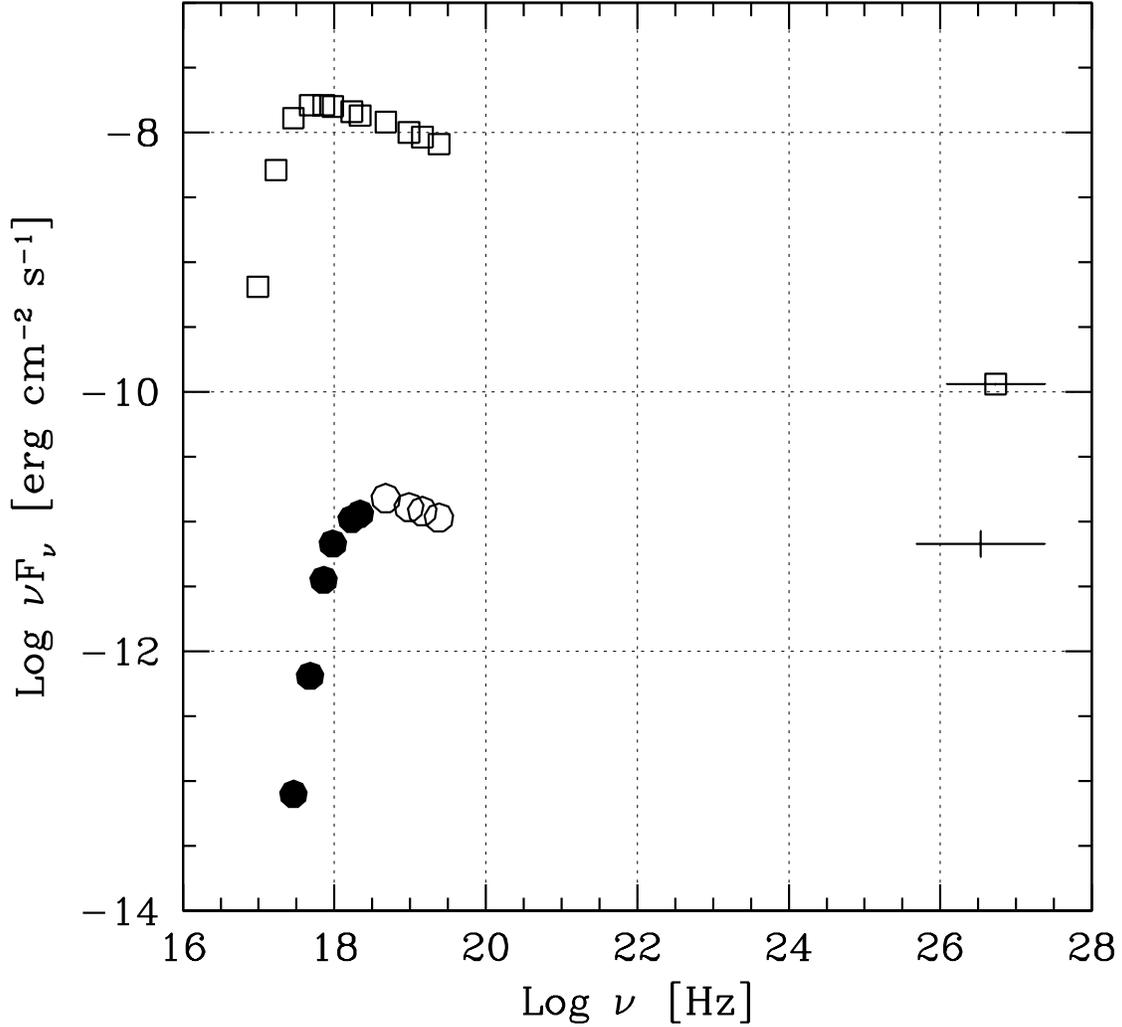} \caption{Spectral energy distribution from radio
to TeV frequencies of IGR J18135-1751/HESS J1713-178 (bottom) and
the Crab Nebula (top). The ASCA X-ray data cover the range  2 to 10 keV,
the INTEGRAL/IBIS ones  correspond to the  20 to 100 keV band and
the HESS point is from 200 GeV to 10 TeV. Flux uncertainties
are within the symbol used.  \label{fig3}}
\end{figure}

The source luminosity is 4 and 3.4 (d/d$_{4}$)$^{2}$ $\times$
10$^{34}$ erg s$^{-1}$ (where d is the distance of the source and
d$_{4}$ = 4kpc) in the 2-10 and 20-100 keV bands, respectively to be
compared to a 0.2-10 TeV luminosity of 1.2-1.9 (d/d$_{4}$)$^{2}$
$\times$ 10$^{34}$ erg s$^{-1}$.  These values are both compatible 
with those observed for a few HESS sources which have been clearly
associated with either shell type or plerionic type supernova
remnants \cite{mal05}. The source Spectral Energy Distribution
(SED) from X-ray to TeV is shown in Fig. \ref{fig3}, together with the
Crab Pulsar Wind Nebula. The shape of the SED resembles that of the
Crab Pulsar Wind Nebula, although with a quite different X-ray/TeV 
gamma-ray ratio. Future XMM-Newton or CHANDRA observations could 
definitively unveil the true nature of this source.\\

\acknowledgments We acknowledge the following funding Agencies: in
Italy, Italian Space Agency financial and programmatic support via
contracts I/R/046/04; in the UK PPARC grant
GR/2002/ 00446.
The Italian authors are grateful to C. Spalletta for the editing of the manuscript and M. Federici for supervising the {\it
INTEGRAL} data analysis System.

We are grateful to Dr. Gustavo Romero for his comments which have improved the scientific quality of the paper.

\clearpage

\clearpage

\end{document}